\documentclass[conference]{IEEEtran}
\IEEEoverridecommandlockouts
\usepackage{cite}
\usepackage{amsmath,amssymb,amsfonts}
\usepackage{algorithm}
\usepackage{algorithmic}
\usepackage{graphicx}
\usepackage{textcomp}
\usepackage{xcolor}
\usepackage{epstopdf}
\def\BibTeX{{\rm B\kern-.05em{\sc i\kern-.025em b}\kern-.08em
    T\kern-.1667em\lower.7ex\hbox{E}\kern-.125emX}}
\begin{document}

\title{Age Based Task Scheduling and Computation Offloading in Mobile-Edge Computing Systems}

\author{
\IEEEauthorblockN{Xianxin Song\IEEEauthorrefmark{1},  Xiaoqi Qin\IEEEauthorrefmark{1}, Yunzheng Tao\IEEEauthorrefmark{2}, Baoling Liu\IEEEauthorrefmark{1}, Ping Zhang\IEEEauthorrefmark{1}, \emph{Fellow, IEEE}}

\IEEEauthorblockA{\IEEEauthorrefmark{1}State Key Laboratory of Networking and Switching Technology, \\
Beijing University of Posts and Telecommunications, Beijing, 100876, China}
\IEEEauthorblockA{\IEEEauthorrefmark{2}Huawei Technologies CO., LTD.}
Email: \{songxianxin, xiaoqiqin, blliu, pzhang\}@bupt.edu.cn, taoyunzheng@huawei.com
}

\maketitle 

\begin{abstract}
To support emerging real-time monitoring and control applications, the timeliness of computation results is of critical importance to mobile-edge computing (MEC) systems. We propose a performance metric called age of task (AoT) based on the concept of age of information (AoI), to evaluate the temporal value of computation tasks. In this paper, we consider a system consisting of a single MEC server and one mobile device running several applications. We study an age minimization problem by jointly considering task scheduling, computation offloading and energy consumption. To solve the problem efficiently, we propose a light-weight task scheduling and computation offloading algorithm. Through performance evaluation, we show that our proposed age-based solution is competitive when compared with traditional strategies.

\end{abstract}

\begin{IEEEkeywords}
mobile-edge computing (MEC), age of task (AoT), task scheduling, computation offloading.
\end{IEEEkeywords}

\section{Introduction}

As the deep integration of mobile Internet and Internet of things, new uses of wireless communication are envisioned in real-time monitoring and control applications such as vehicular networks, industrial control, online facial recognition, etc. The transmission and computation resources in wireless networks are employed to realize networked monitoring and real-time control decision, thus can provide a paradigm shift for wireless communication from data delivery to intelligence acquisition.  In these applications, status updates about the surroundings such as pictures and videos,  need to be processed to reveal the status information embedded in the updates.
However, the limited battery capacity and computation ability of mobile devices restrict the  performance of systems. To address this, mobile-edge computing (MEC) has emerged as a promising technology to solve the contradiction between computation-intensive applications and limited resources of mobile devices \cite{7879258}, \cite{8016573}. By offloading the computation from the mobile device to the MEC server, energy consumption and execution time can be reduced. 

%The exponential growth of smart mobile devices has greatly promoted the development of computation-intensive applications, such as online gaming, face recognition and sensing. However, the limited battery capacity and computation ability of mobile devices restrict the computation-intensive applications. To address this, \emph{mobile-edge computing (MEC)} has emerged as a new technology to solve the contradiction between computation-intensive applications and limited resources of mobile devices \cite{7879258}, \cite{8016573}. By offloading the computation from mobile device to MEC server, energy consumption and execution time can be reduced. Apart from the computation-intensive characteristic, many applications of mobile devices are highly time-sensitive, then it is critical to provide timely information following emergency to support decision making by mobile device. Thus, it is important to evaluate the freshness of information when designing task scheduling. 

Apart from the computation-intensive feature, another key requirement of the emerging real-time monitoring and control applications is the timely situational awareness. That is, the users rely on the computation results to be aware of the surroundings so that right decisions can be made in time. In these applications, stale information is disturbing or even deleterious, therefore the freshness of computation tasks is of critical importance to the performance of MEC systems. The conventional task scheduling and computation offloading strategies are delay-oriented\cite{7874147}, \cite{8371267}. However, every computation task is treated independently and the value of each task does not change over time. Therefore, the  traditional delay-oriented strategies cannot satisfy the requirements of evaluating the freshness of computation tasks and ensure the timeliness of computation results.

Recently, age of information (AoI) is proposed as a  metric to measure the freshness of information \cite{6195689}, \cite{7524524}, \cite{8022894}. The age of a piece of information is defined as the time elapsed since the last received packet is generated at the source.  Previous researches reported that AoI has been further employed to measure the estimation error of collected information in remote estimation systems \cite{8006542}, and effective age of a sample is used to evaluate the time difference between the ideal sample and the actual sample \cite{8406891}. 
%However, AoI cannot be directly applied to MEC systems for the following two reasons. 
%First, for real-time monitoring and control applications, the age value of application should be considered zero, after all tasks of the source are processed in the schedule. There is for the reason that the users have enough understanding of their surroundings after getting the computation results of all tasks. Second, each task reveals requirement of the system for a period of time, which different from each packet just bring the latest system states in status update systems. This brings some changes to the metric of freshness of information.
%For real-time control applications, the age value of application is considered zero, after all tasks of the source are processed in the schedule. There is an underlying reason that the users have enough understanding of their surroundings after getting the computation results of all tasks.

Inspired by AoI, in this paper, we employ the concept of age of task (AoT) to evaluate  the temporal
value of computation tasks that can be revealed through the computation of tasks. In real-time  monitoring and control applications, computation tasks are generated on demand. That is, the generation of tasks is event triggered and captures a change of  status. Therefore, for the sequence of tasks generated by an application, each processed task reveals knowledge of the system only for a short period of time. Similarly, each unprocessed task left in the queue represents that something unknown have happened in the past. Therefore, AoT is defined as the time elapsed since the first unprocessed task left in the queue is generated. It represents the uncertainty of knowledge about the surroundings. To obtain the full awareness of the change of status brought by the passage of time, the computation tasks should be processed in a timely manner.

Task scheduling and computation offloading  problems have been widely studied in MEC systems. In \cite{6574874}, Zhang et al. proposed an energy-efficient binary offloading policy under stochastic wireless channel.
In \cite{7996858}, Zhao et al. proposed a task scheduling and computational resource allocation policy in heterogeneous networks. In \cite{7956189}, Mao et al. proposed an online computation offloading algorithm by  jointly considering  radio and computational resource management  in multi-user MEC systems.
In \cite{8606442}, Tao et al. investigated the optimal computation offloading algorithm when the CPU of helper is opportunistic. 
Different from these existing works, in this paper, we focus on the relationship between the generation time  and the  temporal value of computation tasks. Therefore, we aim to explore an age-based task scheduling and offloading strategy.

The main contributions of this paper can be summarized as follows: 1) Based on the concept of AoI, we employ a new performance metric  called AoT to evaluate the temporal value of computation tasks. 2) We study an age minimization problem   by jointly considering  task scheduling, computation offloading and energy consumption. The formulated problem falls in the form of an integer nonlinear program (INLP). 3) To solve the problem efficiently, we propose a light-weight age-based task scheduling and computation offloading algorithm. 4) We perform numerical studies to show that our proposed solution is competitive when compared with traditional strategies.
%The remainder of this paper is organized as follows. In Section II, we introduce the system architecture of the MEC system, and  present the mathematical modeling and problem formulation of  an energy-constrained AoT minimization problem. In Section III, we propose an efficient task scheduling algorithm. In Section IV, we show the simulation results of different task scheduling algorithms. In Section V, we conclude this paper.

%The remainder of this paper is organized as follows. In Section II, we introduce the system model of the MEC system and propose an new effective age metric, called AoT, to measure the freshness of information in this MEC system. Moreover, we present the mathematical modeling and formulate an energy-constrained AoT minimization problem. In Section III, we propose a greedy-based sub-optimal task scheduling algorithm. In Section IV, we show the simulation results of different task scheduling algorithms. In Section V, we conclude this paper.

The remainder of this paper is organized as follows. In Section II, we present the system architecture for the MEC system, and describe the proposed age minimization problem. In Section III, we propose a light-weight  task scheduling and computation offloading algorithm. In Section IV, we show the simulation results and analyze the results.  Section V concludes this paper.

\section{System Model and Problem Formulation}

\begin{figure}[t]
\centerline{\includegraphics[scale = 0.6]{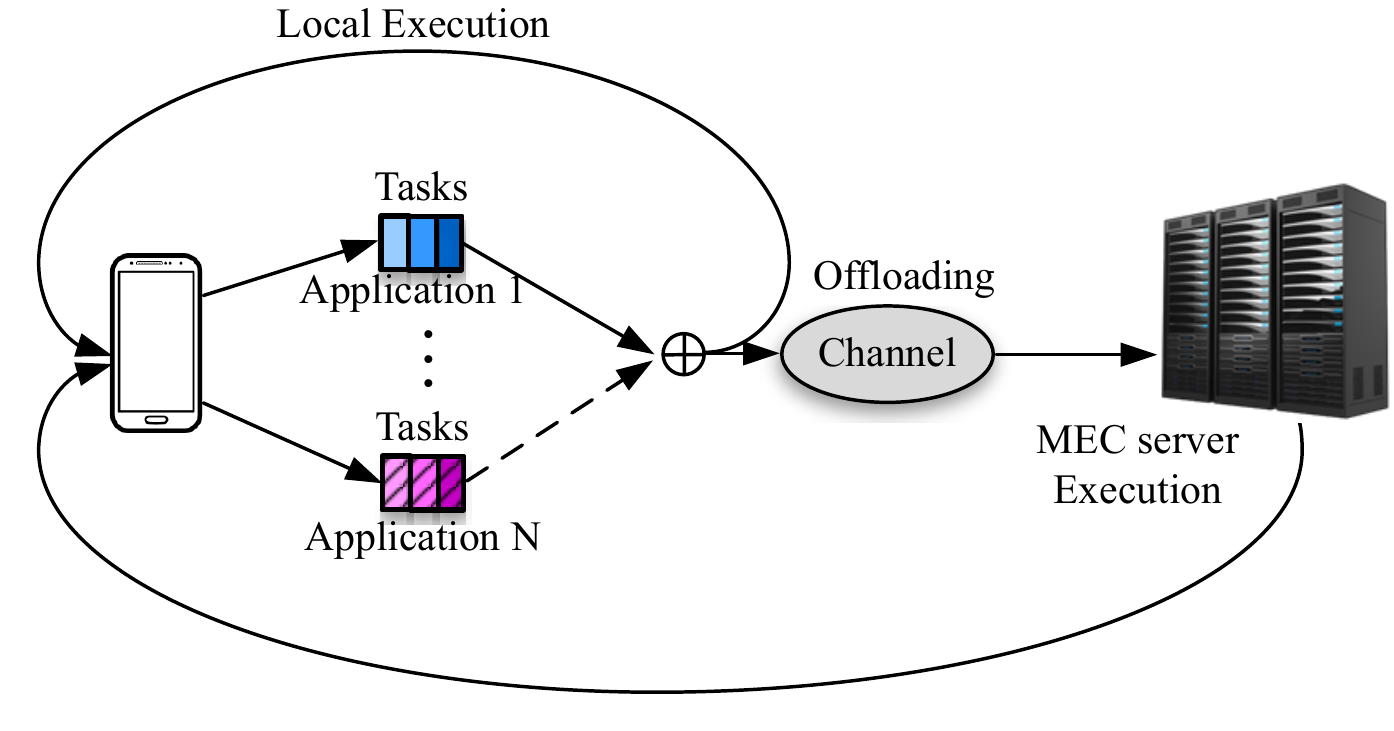}}
\caption{The system model.}
\label{model}
\end{figure}

In this section, we describe a system architecture consisting of a single MEC server and  one mobile device running several applications. We  study an age minimization problem and develop the  mathematical model for task scheduling and computation offloading. 

\subsection{System Architecture}

We consider a status monitoring and control MEC system consisting of a single MEC server and one mobile device as shown in Fig.~\ref{model}. The mobile device runs $N$ applications to monitor several physical phenomena. 
A computation task is generated by application only when a change of status occurs. The status information embedded in the updates can be obtained by processing the tasks. Given a set of tasks for each application, this paper focuses on the task scheduling and computation offloading strategy under energy consumption constraints. 
%Computation task is generated by an application only when the state of source changes. Meanwhile, the samples need to be processed to get the details of status. Therefore, each application has a set of tasks to be processed. In previous work, the generation of tasks is modeled as Poisson process \cite{6189008}. However, in real-time monitoring and control applications, the controller may not be able to predict when a new task will be generated. Therefore, the task scheduling problem in our paper focuses on the tasks that are already in queues.  

\subsection{Mathematical Modeling}

We denote $\mathcal{N}=\{1,2,\ldots,N\}$ as the set of applications with $N=|\mathcal{N}|$ as the total number of applications. We denote $\mathcal{K}_n=\{1,2,\ldots,K_n\}$ as the set of tasks of application $n$ with $K_n=|\mathcal{K}_n|$ as the total number of tasks in each application.  We consider a time slotted system  and the time slot length is $\tau$. We denote $T$ as the total number of time slots. We employ a centralized control architecture, where the MEC server will select one task in each time slot and further partition the data for local computing and computation offloading to the MEC server.

\emph{1) Task Scheduling Constraints:}  
We denote $u_{nk}(t)$ as a binary variable to indicate whether or not task $k$ in application $n$ is selected in time slot $t$.  When the task is selected, $u_{nk}(t)=1$, otherwise, $u_{nk}(t)=0$. In each time slot, at most one task can be selected to be processed. Then we have:
\begin{equation}\label{eq:scheduling 1}
\sum_{n=1}^N\sum_{k=1}^{K_n}u_{nk}(t)\leq1.
\end{equation}

For task scheduling in each application, the processing order of tasks fulfills the first-come-first-served (FCFS) discipline. Task $k+1$ can be selected until the previous task $k$ is completed.
We denote $L_{nk}$ as the data size of  task $k$ in application $n$. Denote $L_{nk}(t)$ as the remaining data size of task $k$ in application $n$ at the beginning time slot $t$. 
Then, when the task is completed, $L_{nk}(t)=0$, otherwise  $L_{nk}(t)>0$. We denote ${\mathcal{K}}'_n$ as the set of tasks of application $n$ except the last task.
Then we have:
\begin{equation}\label{eq:scheduling 2}
u_{n,k+1}(t)\leq1-\frac{L_{nk}(t)}{L_{nk}},\quad n\in\mathcal{N},k\in\mathcal{K}'_n.
\end{equation}
If task $k$ in application $n$ is completed, the value of the terms in the right side of inequalities \eqref{eq:scheduling 2} equals to 1,  otherwise the value is less than 1.

For task scheduling among different applications, if task $k$ is selected to be processed, consecutive time slots will be allocated to this task until it is completed. After completing this task, time slot  will be released for other tasks. When the task is being processed, $0<\frac{L_{nk}(t)}{L_{nk}}<1$,  otherwise  $\frac{L_{nk}(t)}{L_{nk}}=1$ or $\frac{L_{nk}(t)}{L_{nk}}=0$ . Then we have: 
\begin{equation}\label{eq:scheduling 3}
u_{nk}(t)\geq (1-\frac{L_{nk}(t)}{L_{nk}})\frac{L_{nk}(t)}{L_{nk}},\quad n\in\mathcal{N},k\in\mathcal{K}_n.
\end{equation}
If task $k$ in application $n$ is being processed, the value of the terms in the right side of inequalities \eqref{eq:scheduling 3} is a positive number,  otherwise the value is 0.

\emph{2) Computation Offloading Constraints:}
In each time slot, the computation offloading strategy will decide the data size  processed at mobile device, and the data size offloaded to the MEC server.
We denote $D_{loc}(t)$ and $D_{off}(t)$ as the data size processed by local CPU and the MEC server respectively. Obviously, the sum of data size processed in time slot $t$ does not exceed the remaining data size of the selected task. Then we have:
\begin{equation}\label{eq:Dnk(t)}
D_{loc}(t)+D_{off}(t)\leq \sum_{n=1}^N\sum_{k=1}^{K_n} u_{nk}(t)L_{nk}(t).
\end{equation}

Then at the beginning of next time slot, the remaining data size is
\begin{equation}\label{eq:task_length}
L_{nk}(t+1)\!=\!L_{nk}(t)\!-\!u_{nk}(t)(D_{loc}(t)\!+\!D_{off}(t)),n\!\in\!\mathcal{N},k\!\in\!\mathcal{K}_n.
\end{equation}

\emph{3) Energy Consumption Constraints:}
For local computing, the major energy consumption is the operation of CPU.  We denote $\omega$ as the number of CPU cycles required for computing one-bit data at mobile device.  For local computing, CPU operation at a constant CPU-cycle frequency is most energy-efficient. We denote $f(t)$ as the CPU frequency of mobile device in time slot $t$. Then, $f(t)$ can be expressed as
\begin{equation}\label{eq:D_loc(t)}
f(t)=\frac{\omega D_{loc}(t)}{\tau}.
\end{equation}

 In particular, according to the model in \cite{chandrakasan1992low}, the energy consumption per CPU cycle is proportional to the square of the frequency of CPU. The energy consumption per CPU cycle can be modeled by $E_{cyc}=\gamma f^2(t)$, where $\gamma$ is a constant related to chip architecture. Therefore, the energy consumption for  local computing can be given by 
\begin{equation}\label{eq:Energy_loc}
E_{loc}(t)=E_{cyc}\omega D_{loc}(t)=\gamma f^2(t)\omega D_{loc}(t)=\alpha (D_{loc}(t))^3,
\end{equation}
where $\alpha=\gamma \omega^3/\tau^2$.   

For  computation offloading, the major energy consumption of mobile device includes the energy consumption for task computation offloading, and the energy consumption for downloading the computational results of tasks. 
We assume that the computational results are relatively small compared to the input data, therefore the energy consumption for downloading the computational results back to the mobile device is neglected.

%As for computation offloading, we assume that the MEC server provides infinite computation capability compared to mobile device such that the computing time at MEC server can be neglected. Meanwhile, the computational result is relatively small compared to the input data, the time and energy consumption for transmitting the computational result back to the mobile device are neglected. 

We denote $h(t)$ as the channel gain between the mobile device and the MEC server in time slot $t$. Denote $P_{off}(t)$ as the transmitted signal power of mobile device.
Following the empirical model in \cite{6574874},\cite{8606442}, the  transmitted power consumed by reliable transmitting is a convex monomial function with respect to the achievable data rate. 
Then the energy consumption for task computation offloading is
\begin{equation}\label{eq:Energy_off}
E_{off}(t) = P_{off}(t)\tau = \lambda_0\frac{(D_{off}(t)/\tau)^m}{h(t)} \tau= \lambda\frac{(D_{off}(t))^m}{h(t)},  
\end{equation}
where $\lambda=\lambda_0/\tau^{m-1}$. $\lambda_0$ is the energy coefficient incorporating the efforts of bandwidth and noise power. $m$ is the monomial order determined by the modulation-and-coding scheme and takes on values in the typical range of $2\leq m \leq 5$. 
%In this paper, we set the monomial order as $m=3$ \cite{tao2018stochastic}.

We assume that the energy consumed by the mobile device for processing all tasks does not exceed the upper limit $E_{max}$, then we have:
\begin{equation}\label{eq:Energy_sum}
\sum_{t=1}^{T}\{E_{loc}(t)+E_{off}(t)\} \leq E_{max}.
\end{equation}

\emph{4) AoT Constraints:} In real-time monitoring and control applications, users rely on timely situational awareness to make the right decisions.
It is important to obtain timely knowledge of the surroundings, to reduce the uncertainty of the system. The sample strategy of this system is sample-at-change, in which the generation of task is  event triggered and captures a change of status.  
The uncertainty of knowledge about the surroundings appears when a new task is generated, and becomes more severe over time if the task is not processed. When the task is completed, the corresponding knowledge of the specific time interval is revealed by the computation results, and the uncertainty of knowledge about the surroundings  is reduced to the generation time of the first unprocessed task left in the queue. When all tasks have been processed, the uncertainty of knowledge about the surroundings becomes zero.

%In time-sensitive systems, it is important to know the conditions of systems as soon as possible, to reduce the uncertainty of systems. We assume that tasks in applications are generated on demand. New tasks are only generated when the source state change. The uncertainty of sources will appear when new tasks are generated and will grow over time if no task is completed. When a task is completed, the uncertainty of the source will reduce.

%AoI  is commonly used to express the freshness of information in communication systems. However, AoI cannot be directly applied to MEC systems for some reason. First, in MEC systems, tasks are only generated when the states of source changes, which different from the fact that the generation of tasks in status update systems is controllable. It is unreasonable for AoI to continue to grow after all the tasks in the queues have been completed. Second, tasks also reflect the system status after tasks are generated, which different from packets just bring the latest system states in status update systems. This brings some changes to the metric of freshness of information.    

%In this MEC system, when a task has been executed, mobile device will know the situation from the moment when this task was generated to the moment when the next task was generated. And there is no unknown information for the source after all tasks have been executed. To solve this,

\begin{figure}[t]
\centerline{\includegraphics[scale = 0.65]{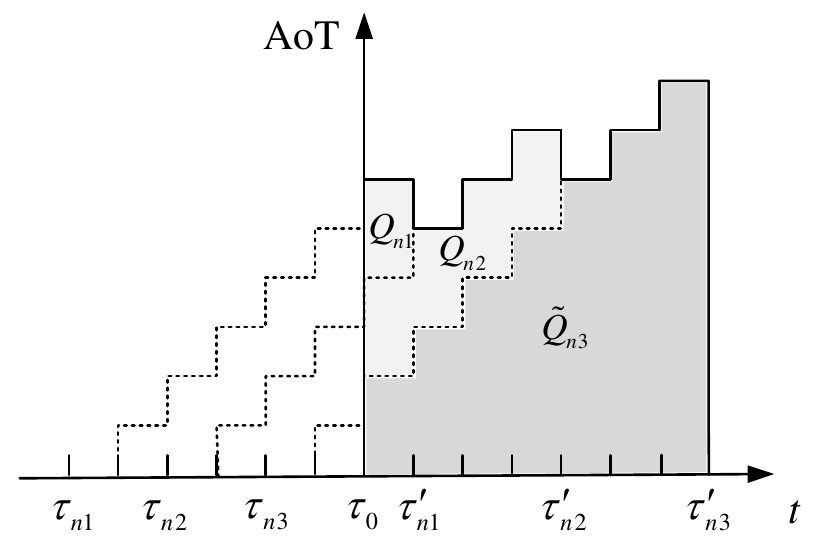}}
\caption{Evolution of AoT in application $n$. The task $k$ in application $n$ is generated at time $\tau_{nk}$ and completed at time $\tau'_{nk}$.}
\label{AoT}
\end{figure}

We employ the concept of AoT to evaluate the temporal value of computation tasks that can be revealed through the computation of a task. Numerically, AoT is the time elapsed since the first unprocessed task left in the queue is generated.  Fig.~\ref{AoT} shows the evolution of AoT in application $n$.
For each application, AoT increases by one per time slot if no task is completed. AoT drops to a smaller value when a task of this application is completed. AoT is reset to zero after all tasks of the application are completed.

As shown in Fig.~\ref{AoT},  we denote $\tau_0$ as the start time of the task scheduling and $a_{n0}$ as the initial age of application $n$. We assume that  task $k$ in application $n$ is generated at time $\tau_{nk}$ and completed at time $\tau'_{nk}$. The initial AoT of application $n$ is the time interval between start time of the task scheduling and generation time of the first task, $a_{n0}=\tau_0-\tau_{n1}$.  Denote $a_n(t)$ as the instantaneous AoT of application $n$ at time slot $t$. It represents the elapsed time since the first unprocessed task left in the queue is generated. Then $a_n(t)$ can be calculated as follows:
\begin{equation}\label{eq:age evolution equation}
   a_n(t) = \begin{cases}
    a_n(t-1)+1, \ \ \;   \text{if no task in application $n$ is } \\
   \qquad \qquad \qquad \quad \text{completed in time slot $(t-1)$,}\\
   t-\tau_{n,k+1},  \qquad \ \text{if task $k$ in application $n$ is}\\
   \qquad \qquad \qquad \quad  \text{completed in time slot $(t-1)$.}\\
    \qquad 0, \qquad \qquad\text{if all tasks in application $n$ are } \\
   \qquad \qquad \qquad \quad \text{completed.}
  \end{cases}
\end{equation}

%We assume there are three tasks generated at $t_{n1}$, $t_{n2}$, $t_{n3}$ respectively.  In the time of the first task waiting to be processed and being processed, AoT in application $n$ increases linearly. After the first task is completed, AoT in application $n$ becomes the age of second task and increases linearly until the second task is completed. After the last task is completed, AoT in application $n$ becomes zero and no longer increases.

%Fig.~\ref{AoT} shows the evolution of AoT in application $n$.

%After all tasks of application $n$ are finished, the age $a_n(t)$ is reset to zero. Therefore, the age function $a_n(t)$ presents finite length sawtooth pattern as shown in Fig.~\ref{AoT}. The value of $a_n(t)$ can be calculated as follows:

Let $A_n$ denotes the overall AoT of application $n$.
The overall AoT is the area under the  finite-length sawtooth ladderlike function in Fig.2, and equals to
\begin{equation}\label{eq:overall aot}
   A_n = \sum_{t=1}^{\tau'_{n,K_n}-\tau_0}a_n(t),\quad n\in\mathcal{N}.  
\end{equation}

As shown in Fig.~\ref{AoT}, the area under this curve can be decomposed into a sum of disjoint geometric parts. The area can be seen as the concatenation of the parallelogram-like area  $Q_{nk}$ for $1\leq\! k \!\leq K_n\!-\!1$ and trapezoid-like area $\tilde{Q}_{n,K_n}$. This decomposition yields
\begin{equation}\label{eq:overall age 1}
   A_n =  \tilde{Q}_{n,K_n}  + \sum_{k=1}^{K_n-1}Q_{nk},\quad n\in\mathcal{N}.  
\end{equation}

We denote $\Delta_{nk}$ as the elapsed time between the generation of tasks $k+1$ and $k$ in application $n$. It follows that
\begin{equation}\label{eq:Delta}
   \Delta_{nk}=\tau_{n,k+1}-\tau_{nk},\quad k\in\mathcal{K}_n^{'}  
\end{equation}
Then the parallelogram-like area $Q_{nk}$ is:  
\begin{equation}\label{eq:Q1 age}
   Q_{nk} = \Delta_{nk}(\tau'_{nk}-\tau_0), \quad n\in\mathcal{N},k\in\mathcal{K}_n^{'},  
\end{equation}
The trapezoid-like area $\tilde{Q}_{n,K_n}$ is:
\begin{equation}\label{eq:Qn age}
\begin{split}
   \tilde{Q}_{n,K_n}&=(\tau_0-\tau_{n,K_n}) + (\tau_0-\tau_{n,K_n} + 1) +\ldots+\\ 
   &(\tau_0-\tau_{n,K_n} + \tau'_{n,K_n}-\tau_0 -1) \\
   &=\frac{(\tau'_{n,K_n}-\tau_0)( \tau_0-2\tau_{n,K_n}+\tau'_{n,K_n} -1)}{2},\quad n\in\mathcal{N},
\end{split}
\end{equation}

%From Fig.~\ref{AoT}, we can find that the area $Q_{nk}$ for $1\leq\! k \!\leq K_n\!-\!1$ can be calculated as the sum of multiple equal rectangular areas. The area of a single rectangle is related to the task generation interval and the number of rectangles is related to the completion time of task. The area $\tilde{Q}_{n,K_n}$ can be calculated as the sum of multiple uniformly enlarged rectangular areas. It is related to the generation time and completion time of task $K_n$.

Therefore, the overall AoT of application $n$ can be calculated as follows:
\begin{equation}\label{eq:overall age}
\begin{split}
   A_n = &\frac{(\tau'_{n,K_n}-\tau_0)^2}{2}+(\tau _0-\tau_{n,K_n}-\frac{1}{2})(\tau'_{n,K_n}-\tau_0)+\\
   &\sum_{k=1}^{K_n-1}\Delta_{nk}(\tau'_{nk}-\tau_0),\quad n\in\mathcal{N}. 
\end{split}
\end{equation}

\subsection{Problem Formulation}

To ensure the timely situational awareness about the surroundings, we aim to minimize the sum AoT of all applications under energy consumption constraints. The problem can be formulated as follows:
%To ensure the timely situational awareness, we aim to minimize the sum age of all applications under the total energy consumption constraint. At the starting time, the MEC server makes scheduling decision based on the time stamps and length  of tasks and the channel gains in the following. The problem is formulated as
\begin{align}\label{eq:OPT1}\notag 
\textbf{(P1)} \qquad & \min_{u_{nk}(t),D_{loc}(t),D_{off}(t)} \  \sum_{n=1}^{N}A_n \\\notag 
\text{s.t.} \qquad & \textrm{Task scheduling constraints: }\eqref{eq:scheduling 1} - \eqref{eq:scheduling 3};\notag \\
 &\textrm{Computation offloading constraints: }\eqref{eq:Dnk(t)}, \eqref{eq:task_length};\notag \\
 &\textrm{Energy consumption constraints:}\eqref{eq:Energy_loc} - \eqref{eq:Energy_sum}\notag \\
&\textrm{AoT constraints: }\eqref{eq:overall age 1}, \eqref{eq:Delta},\eqref{eq:overall age}.\notag 
\end{align}

In this formulation, $D_{loc}(t)$ and $D_{off}(t)$ are integer variables. $u_{nk}(t)$ is binary variable. $N$, $K_n$, $L_{nk}$, $\tau_{nk}$, $h(t)$  are constants. The formulated problem is an INLP, which is intractable. To remove the nonlinear terms in constraint \eqref{eq:Energy_loc} and \eqref{eq:Energy_off}, we employ the piecewise linear approximation technique. Then the problem is transformed into an integer linear program, which can be solved by a commercial solver such as Gurobi \cite{gurobi}.

\begin{algorithm}[t]
\caption{ An efficient task scheduling algorithm} 
\label{order} 
\renewcommand{\algorithmicrequire}{\textbf{Input:}} 
\renewcommand{\algorithmicensure}{\textbf{Output:}}
  \begin{algorithmic}[1]
    \STATE $empty(n)\!\leftarrow\!0$, $num(n)\!\leftarrow\!1,  order\! \leftarrow \!\emptyset, i \!\leftarrow \!1, n\!\in\!\mathcal{N}$ 
    \WHILE{$sum(empty) \leq N -1$}
      \FOR{$n\in\mathcal{N}$}
        \STATE Compute $\delta_{n}$
      \ENDFOR
      \STATE $c\leftarrow argmax(\delta_{n})$, $order(i)\leftarrow c$, $i\leftarrow i + 1$, $E(i) \leftarrow E_{c,num(c)}$
      \STATE  Perform task computation offloading, update AoT of each application.
      \STATE $E(i) \leftarrow E'_{c,num(c)}$
      \IF{$num(c)\leq K_n -1$}
        \STATE $num(c) \leftarrow num(c) + 1$
      \ELSE
        \STATE $empty(c) \leftarrow 1$
      \ENDIF
    \ENDWHILE
    \FOR{$j = 1 \ to \ \sum_{n=1}^{N}K_n$}
    \STATE $E(j)=E_{max}-sum(E)+E(j)$
    \STATE $empty(n)\leftarrow0$, $num(n)\leftarrow 1, i \leftarrow 1, n\in\mathcal{N} $ 
        \WHILE{$sum(empty) \leq N -1$}
      \STATE $c\leftarrow order(i), E_{c,num(c)}\leftarrow E(i)$, $i\leftarrow i + 1$
      \STATE repeat lines 7 - 13 
    \ENDWHILE
     \ENDFOR
  \end{algorithmic}
\end{algorithm}

\section{An Efficient Solution}

In this section, we present a light-weight age-based task scheduling and computation offloading algorithm. The basic idea is as follows. For task scheduling, we should select the task which reveals the most knowledge about the surroundings and least waiting time for other tasks. For computation offloading, we should decide the data size processed in each time slot as well as the exact portions for local computing and computation offloading to minimize the processing time of the selected task under energy  constraints.

For task scheduling, if a task in application $n$ is selected for processing, the change of sum AoT of all applications consists of two parts: 1) the age reduction of application $n$ caused by completing the task, 2) the age increment of applications caused by waiting for the processing of the task. We denote $\delta_{n}$ as the change of the sum AoT of all applications after completing a task in application $n$. Denote $\delta_{n}^{r}$ as the age reduction by completing the task and $\delta_{n}^{i}$ as the age increment due to waiting for the processing of the task, then we have $\delta_{n}=\delta_{n}^{r}-\delta_{n}^{i}$.

%different length of  tasks results in different task processing time.

%Equation \eqref{eq:age evolution equation} shows that age of task in application $n$ decreases from $a_n(t-1)+1$ to $t-t_{n,k+1}$ once task $k$ of application $n$ is completed in time slot $t$. The age reduction caused by  processing  task $k$ in application $n$ equals the task generation interval $\Delta_{nk}$. The greater the task generation interval, the greater the age reduction after completing the task. Meanwhile, the generation interval of tasks indicates the duration of the state. The greater the generation interval from this task to the next task, the longer the duration of the state. Because we want to obtain knowledge of the surroundings in each time slot, the amount of valid information contained in each task is proportional to the the task generation interval. Intuitively, the task with large the task generation interval should be prioritized. AoT of each application increases during task processing. Different length of  tasks results in different task processing time.

%We denote $\delta_{n}$ as the change of the sum age of all applications after completing a task in application $n$. $\delta_{n}$ consists of two terms: the age reduction $\delta_{n}^{r}$ of this application due to completing this task and the age increase $\delta_{n}^{i}$ of applications with unprocessed task during processing, then we have:

First, we look into the details of age reduction caused by completing task $k$ in application $n$. For the set of tasks except the last task, $\delta_{n}^{r}$ equals to $\Delta_{nk}$, which is a constant determined by the task generation interval times defined in constraint \eqref{eq:Delta}. For the last task,  $\delta_{n}^{r}$ equals to $a_{n}(\tau'_{n,K_n}-\tau_0)+1$ because AoT is reset to zero after complete this task, where $a_{n}(\tau'_{n,K_n}-\tau_0)$ is the AoT of application $n$ at time slot $\tau'_{n,K_n}-\tau_0$. Then, we consider the age increment of applications. We denote $N'$ as the number of applications that have unprocessed task and $s_{nk}$ as the number of time slots needed to process task $k$ in application $n$.  Then  the total age increment caused by waiting for the processing of task $n$ in application $n$ is $s_{nk}N'$. 

For computation offloading, we aim at finding the optimal data size for local computing and offloading to minimize the processing time under energy constraints. We denote $E_{nk}$ as the energy assigned to task $k$ in application $n$.  We assign an initial energy to each task, which is proportional to the cube of the task length. Then we have:
\begin{equation}\label{eq:}
   E_{nk}=\frac{(L_{nk})^3}{\sum_{n=1}^N\sum_{k=1}^{K_n} (L_{nk})^3}E_{max}. 
\end{equation}
We denoet $t^s_{nk}$ as the time when task $k$ in application $n$ starts to be processed.
The computation offloading problem is formulated as follows:
\begin{align}\label{eq:OPT2} \notag
\textbf{(P2)} \qquad & \min_{D_{loc}(t),D_{off}(t)} \  s_{nk} \\
\text{s.t.} \qquad & \eqref{eq:Energy_loc}, \eqref{eq:Energy_off},\notag \\
&\sum_{t=t^s_{nk}}^{t^s_{nk}+s_{nk}}\{D_{loc}(t)+D_{off}(t)\}=L_{nk}, \notag \\
&\sum_{t=t^s_{nk}}^{t^s_{nk}+s_{nk}}\{E_{loc}(t)+E_{off}(t)\}\leq E_{nk}.\notag 
\end{align}

The computation offloading problem includes  two parts: 
How many bits can be processed in each time slot? Within each time slot, how many bits should be processed locally and offloaded to MEC server, respectively?
If the data size  can be processed in time slot $t$  is fixed, we aim to find an offloading scheme to minimize the energy consumption based on channel condition. We denote $D(t)$ as the data size processed in time slot $t$.
%the data partition for processing in different time slots and the data partition for local computing and computation offloading in a single time slot. As for the data partition in a single time slot, the objective is to minimize the energy consumption according to channel gains. We denote $D(t)$ as the sum data size processed in time slot $t$ by local CPU and MEC server. Then we have $D_{off}(t)=D(t)-D_{loc}(t)$.  
The energy consumption in a single time slot is 
\begin{equation}\label{eq:Energy_slot}
   E(t)=\alpha (D_{loc}(t))^3+\lambda\frac{(D(t)-D_{loc}(t))^m}{h(t)}. 
\end{equation}

Notably, Equation \eqref{eq:Energy_slot} is  a convex function. We set the monomial order $m$ as 3 \cite{8606442}.
Using the Lagrangian method, the optimal data allocation for local execution and MEC server execution, $D_{loc}(t)$ and $D_{off}(t)$, are given by
\begin{equation}\label{eq:allocate_data}
D_{loc}(t)=\frac{D(t)}{1+\sqrt{\frac{\alpha h(t)}{\lambda}}}\ \text{and} \ D_{off}(t)=\frac{D(t)}{1+\sqrt{\frac{\lambda}{\alpha h(t)}}}.
\end{equation}
Then the minimum energy consumption for execute $D(t)$ bits data in a single time slot $t$ can be obtained as
\begin{equation}\label{eq:allocate_data_off}
E(t)=\alpha(D(t))^3(1+\sqrt{\frac{\alpha h(t)}{\lambda}})^{-2}. 
\end{equation}

To obtain the value of $D(t)$, we apply feasibility test to find the minimum number of time slots needed to process a task under energy constraints. We first fix the task completion time as one time slot and $D(t)=L_{nk}$ to calculate the energy needed based on Equation \eqref{eq:allocate_data_off}. If the obtained energy $E'_{nk}$ is smaller than the energy limit ($E_{nk}$), the value of $D(t)$ is feasible. Otherwise, we increase the task completion time by one slot and   calculate the date sizes processed in each time slot interactively until the resulting energy consumption meet the energy consumption constraint. Note that the data partition among different time slots can be obtained by using Lagrange method for Equation \eqref{eq:allocate_data_off}.
\begin{equation}\label{eq:data}
\frac{D(t_1)}{D(t_2)}=\frac{1+\sqrt{\frac{\alpha h(t_1)}{\lambda}}}{1+\sqrt{\frac{\alpha h(t_2)}{\lambda}}}.
\end{equation}

After determining the processing order of tasks, if there is energy left, we assign the rest energy to previous tasks to further reduce the age increment caused by task waiting.

%Since the completion time of the task is an integer number of time slots, the actual energy consumption of each task is no more than the energy assigned to this task. After determining the processing order of tasks, we assign the rest energy to previous task to reducing the waiting time of all tasks. 

%Equation \eqref{eq:allocate_data} and Equation \eqref{eq:data} show the proportional offloading determined by the channel state $h(t)$. When the channel status between mobile device and MEC server is good, more data will be offloaded to MEC server to process to reduce energy consumption. And between different time slots, more data is processed in time slots with good channel gains. 

\section{Performance Evaluation}
In this section, we present simulation results to first answer the following question: Is there a difference between age-based task scheduling and traditional delay-based task scheduling? 
Moreover, we evaluate the performance of our proposed algorithm and compare the performance between age-optimal strategy and delay-optimal strategy. As for delay-optimal strategy, the completion time of all tasks is minimized. As a performance benchmark, we also simulate the MEC-only strategy where each task is offloaded to MEC server for computing using round-robin task scheduling. Note that the optimal solutions in these three strategies are obtained using a commercial solver Gurobi. 
%Moreover, we also compare the solution obtained by our proposed light-weight algorithm with the optimal solutions

We show that age-optimal strategy outperforms delay-optimal strategy in minimizing the sum AoT of all applications, while it also offers competitive solution in minimizing completion time of all tasks when compared with delay-optimal strategy. We also notice that the MEC-only strategy offers poor performance in terms of both sum age and completion time of tasks.  This shows that judicious design of task scheduling and computation offloading strategy is critical.

\subsection{Simulation Settings}
%The simulation parameters are set as follows unless specified otherwise. 
We consider the mobile device runs 3 applications and each application has 3 tasks that are randomly generated. The time stamp of each task follows the uniform distribution with $[1,8]$.  The starting time of the scheduling horizon is $\tau_0=10$.
The length of time slot is set as 10 ms. The data size of each task follows the uniform distribution with $[400,600]$ (bits). For local computing, the effective energy coefficient of the CPU at mobile device is set as $\gamma=10^{-28}$. Computing one-bit data at mobile device require $\omega=10^5$ CPU cycles. For MEC computation offloading, the channel gain in each time slot follows the uniform distribution with $[10^{-5},10^{-3}]$. The energy coefficient and the monomial order are set as $\lambda_0=10^{-17}$ and $m=3$, respectively.  For each comparison study, we generate 
50 random instances and take the average from the results.

%For performance comparison, we consider Min AoI-Optimal Solution, Min Delay and MEC-only policies. The objective of minimizing delay is reducing the completion time of all tasks. The MEC-only policy offloads all tasks to MEC server for computing and the task selection strategy is round-robin. The performances of these policies are obtained by Gurobi. For comparison study, we generate 20 random instances and take the average from the results. 

\subsection{Results}

%There are mainly two indicators for testing. The first indicator is the sum of age of all applications, indicating the freshness of information in this MEC system. The second indicator is the completion time of all tasks of all applications, which is a performance indicator in traditional MEC systems. The completion time of total tasks here is the total number of time slots spends on completing all tasks of all applications.

%For performance comparison, we consider Min AoI-Optimal Solution, Min Delay and MEC-only policies. The objective of minimizing delay is reducing the completion time of all tasks. The MEC-only policy offloads all tasks to MEC server for computing and the task selection strategy is round-robin. The performances of these policies are obtained using Gurobi, respectively. We also calculated the performance of our proposed algorithm. 

%The curves of the sum age of applications  versus the total energy consumption are plotted in Fig.~\ref{simulation_aot}. The curves of completion time of all tasks versus the total energy consumption re plotted in Fig.~\ref{delay}.

%Fig.~\ref{simulation_aot} and Fig.~\ref{delay} show  the trend of the sum age of applications and the completion time of all tasks obtained by the proposed algorithm, age-optimal strategy, delay-optimal strategy, and MEC-only strategy as the total energy consumption increases from 0.12 J to 0.18 J, respectively.

\begin{figure}[t]
\centerline{\includegraphics[scale = 0.48 ]{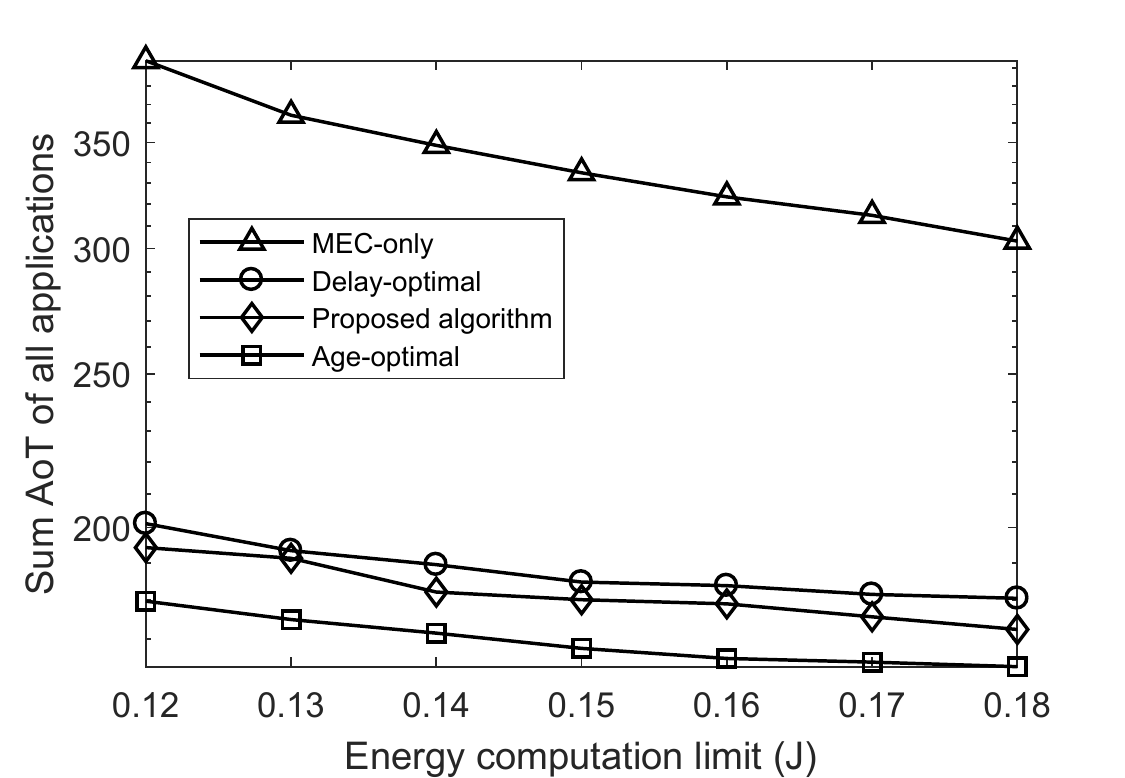}}
\caption{Sum AoT of all applications when energy consumption limit increases from 0.12 (J) to 0.18 (J).}
\label{simulation_aot}
\end{figure}
\begin{figure}[t]
\centerline{\includegraphics[scale = 0.48]{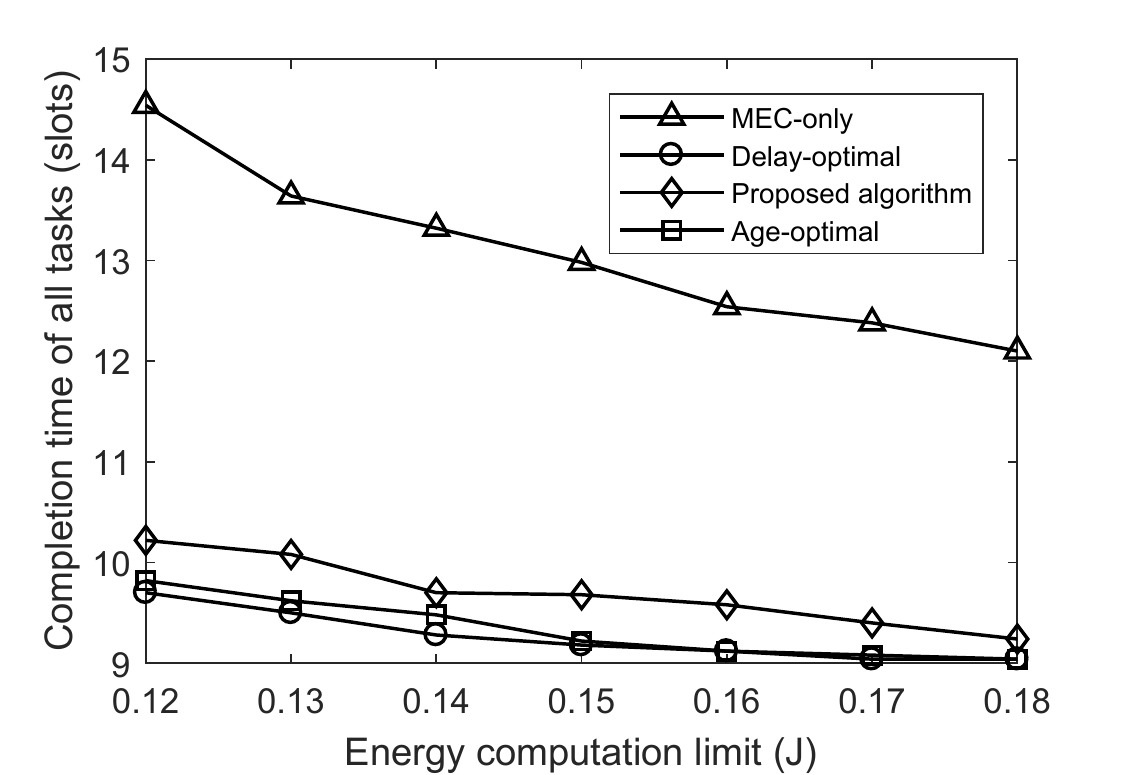}}
\caption{Completion time of all tasks when energy consumption limit increases from 0.12 (J) to 0.18 (J).}
\label{delay}
\end{figure}

We first compare the achievable sum AoT of all applications obtained by the four strategies for different energy consumption limits.   Fig.~\ref{simulation_aot} shows the trend of sum AoT of all applications as the energy consumption limit  increases from 0.12 J to 0.18 J.  Each point on the curve is averaged over results from 50 randomly generated instances.
As shown in Fig.~\ref{simulation_aot},
the objective values obtained by the four strategies decrease as the energy consumption limit increases, as expected.
It shows that (i) MEC-only strategy yields the worst performance; (ii) the age-optimal strategy and the proposed algorithm outperform the delay-optimal strategy (with an average gap of 18.2  slots and 5.7  slots, respectively); (iii) the average ratio between the objective values obtained by the proposed algorithm and those from Gurobi is 93.2\%.

%Due to lack of efficient task scheduling and computation  offloading, the objective values obtained by the MEC-only strategy are worst compared to other strategies.The objective values obtained by the delay-optimal strategy is 18.19 (time slots) larger than those by age-optimal strategy. The average ratio between the objective values obtained by age-optimal strategy and those by the proposed strategy is 93.2\%. 
%It is easy to observe that (i) Age-optimal strategy performs better in processing tasks in a timely manner when compared with delay-optimal strategy. (ii) Age-optimal strategy are competitive in minimizing completion time of all tasks when compared with delay-optimal strategy.

We also compare the completion time of all tasks obtained by the four strategies for different energy consumption limits. 
Fig.~\ref{delay} shows the trend of completion time of all tasks as the energy consumption limit increases from 0.12 J to 0.18 J. Each point on the curve is averaged over results from 50 randomly generated instances. As shown in Fig.~\ref{delay}, the completion time needed by the four strategies decreases as the energy consumption limit increases, as expected.
It shows that (i) MEC-only strategy yields the worst performance; (ii) the performance of both age-optimal strategy and the proposed algorithm are competitive when compared with the delay-optimal strategy (with an average gap of 0.07 slots and 0.43 slots, respectively); (iii) the average ratio between the objective values obtained by the proposed algorithm and those by delay-optimal strategy is 95.6\%.

\section{Conclusions}

The emerging real-time monitoring and control applications require timely situational awareness. Therefore, evaluating the temporal value of computation tasks is urgently needed when designing task scheduling strategy.  In this paper, we employed the concept of AoT to evaluate the temporal value of information. We studied an age minimization problem in a MEC system consisting of a single MEC server and one mobile device running several applications. The problem formulation involves  task scheduling constraints, computation offloading constraints, energy consumption constraints, and falls in the form of an INLP. To solve the problem efficiently, we proposed a light-weight task scheduling and computation offloading algorithm. Simulation results showed that the performance of our proposed age-based scheduling strategy is competitive when compared with traditional strategies.

\section*{Acknowledgment}

This paper is supported by the National Science Foundation for Young Scientists of China Project No.042700349, the Beijing University of Posts and Telecommunications Project No. 500418759, and the State Key Laboratory of Networking and Switching Technology Project No. 600118124.
\renewcommand\refname{Reference}
\bibliographystyle{ieeetr}
\bibliography{reference}

\end{document}